\documentclass[12pt]{article}
\newcommand{\mapright}[2]
{\mathop{\hbox to 8mm{\rightarrowfill}}
\limits^{\scriptstyle #1}_{\scriptstyle #2}}

\makeatletter
\@addtoreset{equation}{section}
\makeatother

\begin{document} %
\begin{flushright}
{KOBE-TH-02-05}\\
{DIAS-STP-03-01}\\
\end{flushright}
\vspace{5mm}
\vspace*{6mm}
\begin{center}
{\Large \bf
Supersymmetry in quantum mechanics with\\
point interactions}
\vspace{10mm} \\
Tomoaki Nagasawa\footnote{e-mail:
nagasawa@phys.sci.kobe-u.ac.jp (T. Nagasawa)},
Makoto Sakamoto\footnote{e-mail:
sakamoto@phys.sci.kobe-u.ac.jp (M. Sakamoto)},
Kazunori Takenaga\footnote{e-mail:
takenaga@synge.stp.dias.ie (K. Takenaga)}
\vspace*{10mm} \\
{\small \it
${}^1 $Graduate School of Science and Technology, Kobe University,
Rokkodai, Nada, \\ Kobe 657-8501, Japan
\\
${}^2 $Department of Physics, Kobe University,
Rokkodai, Nada, Kobe 657-8501, Japan
\\
${}^3 $School of Theoretical Physics,Dublin Institute for Advanced
Study,
\\ 10 Burlington Road,Dublin 4,Ireland
}
\vspace{10mm}
\\
Abstract
\vspace{4mm}
\\
\begin{minipage}[t]{130mm}
\baselineskip 6mm 
We investigate supersymmetry in one-dimensional quantum mechanics with
point interactions.
We clarify a class of point interactions compatible with supersymmetry
and present $N=2$ supersymmetric models on a circle with two point
interactions as well as a superpotential.
A hidden $su(2)$ structure
inherent in the system
plays a crucial role to construct the $N=2$ supercharges. 
Spontaneous supersymmetry breaking due to point interactions 
and an extension to
higher $N$-extended supersymmetry are also discussed.
\end{minipage}
\end{center}
\vspace{8mm}
\noindent
%
\newpage
\baselineskip 6mm 

\section{Introduction}
Quantum mechanics in one dimension admits point singularities as
interactions of zero range.
A point interaction is parameterized by the group
$U(2)$ \cite{u(2)1,u(2)2,u(2)3}, and the varieties
of connection conditions between a wavefunction and
its derivative due to $U(2)$
lead to various interesting physical phenomena, such as
duality \cite{duality1,duality2},
the Berry phase \cite{berry1,berry2} and scale anomaly \cite{symmetry}.
Furthermore, for some specific choices of point interactions, there
occurs a double
degeneracy in the energy level, which suggests the existence of
supersymmetry
\cite{symmetry}.


The purpose of this paper is to examine supersymmetry in 
quantum mechanics on a circle with not only point interactions
but also potentials regular 
except for point singularities
and to give a brief discussion on $N$-extended
supersymmetry with $N>2$.
We first clarify a class of connection conditions compatible 
with supersymmetry.
We then find $N=2$ supersymmetric models, which differ from
the usual supersymmetric Witten model \cite{witten}
in several ways.
Our system consists of only one bosonic degree of freedom,
and \lq\lq bosonic" and \lq\lq fermionic" states are assigned
according to eigenvalues of a parity-like transformation.
The derivative of a superpotential $W'(x)$ is restricted to
be parity-odd, while there is no such restriction in the
Witten model.
Further, $W'(x)$ is allowed to have discontinuity at singularities
in our models.
Our system naturally possesses some discrete transformations that
form an $su(2)$ algebra and they are used in constructing 
$N=2$ supercharges in place of the Pauli matrices in the
Witten model.
Spontaneous supersymmetry breaking occurs in the Witten model
if zero energy solutions are not normalizable.
On the other hand, in our models, it occurs if zero energy
solutions are incompatible with connection conditions
at point interactions.

\section{Quantum mechanics with point interactions}
In this section, we give a brief review of one-dimensional quantum
mechanics with
point interactions and present a setup of our model.

In this paper, we consider quantum mechanics on a circle
$S^1 (-l<x\le l)$ on which two
point interactions are put at $x=0$ and $x=l$.
Although our analyses can equally apply for a noncompact space,
characteristic features of
point interactions will become more apparent 
in our model, namely, in discussing
spontaneous breaking of supersymmetry.
A point interaction is specified by a characteristic matrix $U\in
U(2)$, and a wavefunction
$\varphi(x)$ and its derivative are required to obey the 
connection condition at, say
$x=0$ \cite{cc,symmetry}
\begin{equation}
    (U-{\bf 1})\Phi + iL_0 (U+{\bf 1})\Phi' =0,
    \label{cc}
\end{equation}
where $L_0$ is an arbitrary nonzero constant and
\begin{equation}
    \Phi \equiv
    \left(
        \begin{array}{c}
            \varphi(0_+) \\
            \varphi(0_-)
        \end{array}
    \right)
    ,~~~
    \Phi'  \equiv
    \left(
            \begin{array}{c}
            \varphi'(0_+) \\
            -\varphi'(0_-)
        \end{array}
    \right).
\end{equation}
Here, $0_\pm$ denote $0 \pm \varepsilon$ with an infinitesimal positive
constant
$\varepsilon$, and $\varphi'(x)=\frac{d\varphi(x)}{dx}$. The connection
condition
at $x=l$ is similarly specified with a (generally different)
characteristic matrix
$\bar{U} \in U(2)$.

For later convenience, let us rewrite the connection condition
$(\ref{cc})$ to make a
supersymmetric structure clearer. To this end, we first note that since
the four
generators of $U(2)$ are taken to be $\{ {\bf 1},\vec{\sigma}\}$, any
$U(2)$
matrix can be parameterized as
\begin{equation}
    U_{g}(\theta_+,\theta_-) =
    \exp\left\{i\theta_+ P_g^+ +i\theta_- P_g^-\right\},
\label{theta}
\end{equation}
where
\begin{eqnarray}
    P_g^{\pm}&=&\frac{1}{2}( {\bf 1} \pm g),\\
     g &=& \vec{\alpha} \cdot \vec{\sigma} \quad {\rm with} \
      (\vec{\alpha})^2=1.
\end{eqnarray}
The $P_g^{\pm}$ can be regarded as projection matrices,
\begin{equation}
    (P_g^{\pm})^2 = P_g^{\pm},~~P_g^{\pm}P_g^{\mp}
    = 0,~~P_g^+ +P_g^-={\bf 1}.
\end{equation}

Let us next introduce three discrete transformations. The first is the
parity transformation
$\cal P$ defined by
\begin{equation}
    {\cal P}:\ \ \varphi(x) \rightarrow
             {\cal P} \varphi(x) = \varphi(-x) .
\end{equation}
The half-reflection transformation ${\cal R}$ is inherent in quantum
mechanics with point
singularities and is defined by
\begin{equation}
    {\cal R} :\ \  \varphi(x) \rightarrow {\cal R}
              \varphi(x) = (\Theta(x)-\Theta(-x))\varphi(x),
\end{equation}
where $\Theta(x)$ is the Heaviside step function.
The third transformation $\cal{Q}$
is defined by $ {\cal Q}\equiv -i {\cal R P}$.
An important observation is that the set
$\{ {\cal P}_1={\cal P},{\cal P}_2={\cal Q},{\cal P}_3={\cal R} \} $
forms the $su(2)$ algebra of spin $1\over2$, i.e.
\begin{eqnarray}
    [{\cal P}_i,{\cal P}_j]      &=& 2 i \sum_{k=1}^{3}
      \epsilon_{ijk} {\cal P}_k , \\
    \{ {\cal P}_i , {\cal P}_j\} &=& 2 \delta_{ij}.
\end{eqnarray}
It turns out that they are essential ingredients to construct $N=2$
supercharges in our formulation,
as we will see later.
By use of $ {\cal P}_{j}$, we can introduce an
$su(2)$ element ${\cal G}$
associated with $g=\vec{\alpha}\cdot \vec{\sigma}$ as
\begin{equation}
    {\cal G} \equiv \vec{\alpha}\cdot \vec{ {\cal P}}.
\end{equation}
Since $ {\cal G}^2=1$, we can decompose any wavefunction $\varphi(x)$
into two
eigenfunctions $\varphi_{\pm}(x) \equiv  \frac{1}{2} (1\pm {\cal G})
\varphi(x)$ satisfying
\begin{equation}
    {\cal G} \varphi_{\pm}(x) = \pm \varphi_{\pm}(x).
\end{equation}
It is not difficult to show that the connection condition (\ref{cc}) at
$x=0$ splits into
\begin{eqnarray}
    \sin{\frac{\theta_+}{2}}\ \varphi_+(0_+)+ L_0
      \cos{\frac{\theta_+}{2}}\;\varphi'_+(0_+) &=& 0, \nonumber \\
    \sin{\frac{\theta_-}{2}}\;\varphi_-(0_-)- L_0
      \cos{\frac{\theta_-}{2}}\;\varphi'_-(0_-) &=& 0.
\label{cc0}
\end{eqnarray}
The connection condition at $x=l$ is assumed to be specified by the
characteristic matrix
$U_g (\bar{\theta}_+,\bar{\theta}_-)$, i.e.
\begin{eqnarray}
    \sin{\frac{\bar{\theta}_+}{2}}\;\varphi_+(l)+ L_0
      \cos{\frac{\bar{\theta}_+}{2}}\;\varphi'_+(l) &=& 0, \nonumber \\
    \sin{\frac{\bar{\theta}_-}{2}}\;\varphi_-(-l)- L_0
      \cos{\frac{\bar{\theta}_-}{2}}\;\varphi'_-(-l) &=& 0,
\label{ccl}
\end{eqnarray}
where $(\bar{\theta}_+,\bar{\theta}_-)$ are, in general, different
from $(\theta_+,\theta_-)$.
\section{Compatibility with supersymmetry}
In this section, we show that the connection conditions
$(\ref{cc0})$ and $(\ref{ccl})$ are,
in general, inconsistent with supersymmetry transformations, and
clarify how compatibility with
supersymmetry restricts the values of $(\theta_+,\theta_-)$ and
$(\bar{\theta}_+,\bar{\theta}_-)$ in eqs.$(\ref{cc0})$
and $(\ref{ccl})$.

Let us first discuss the quantum system only with point interactions,
so that the Hamiltonian
is given by
\begin{equation}
    H=-\frac{1}{2} \frac{d^2}{dx^2}
\end{equation}
except for the singular points.
Here we have set $\hbar =1$ and the mass $m=1$ for simplicity.
An extension to models with
potential terms will be given in Section 5.
If the system is supersymmetric, the Hamiltonian will be written, in
terms of a supercharge $Q$,
as $H=2 Q^2$.
It follows that the supercharge is expected to be proportional to the
derivative, $Q \propto \frac{d}{dx}$.
This, however, causes a trouble to construct supercharges because
$\varphi'(x)$ does not,
in general, obey the same connection condition as $\varphi(x)$, and
hence the state $Q\varphi(x)$ would not belong to the Hilbert space of
the model.

To find a class of connection conditions compatible with supersymmetry,
let us examine a
supersymmetric partner $\chi(x) \equiv Q \varphi(x)$ of any state
$\varphi(x)$ that satisfies the connection condition
$(\ref{cc})$ and the Schr{\"o}dinger equation
\begin{equation}
    H \varphi(x) =E \varphi(x).
    \label{Seq}
\end{equation}
Since the supercharge is proportional to $\frac{d}{dx}$,
$\chi(0_{\pm})$ will be given, in general, by a linear combination of
$\varphi'(0_+)$ and $\varphi'(0_-)$ such as
\begin{equation}
    \Phi_{\chi} \equiv
    \left(
        \begin{array}{c}
            \chi(0_+) \\
            \chi(0_-)
        \end{array}
    \right) =M
    \left(
        \begin{array}{c}
            \varphi'(0_+) \\
            -\varphi'(0_-)
        \end{array}
    \right)
\label{chi}
\end{equation}
for some invertible constant matrix $M$. Since $\varphi(x)$
satisfies the Schr{\"o}dinger
equation $(\ref{Seq})$, $\varphi''(x)$ is proportional
to $\varphi(x)$. 
This fact implies that
$\chi '(0_{\pm})$ should be related to $\varphi(0_{\pm})$ as
\begin{equation}
    \Phi'_{\chi} \equiv
    \left(
        \begin{array}{c}
            \chi'(0_+) \\
            -\chi'(0_-)
        \end{array}
    \right) =E \tilde{M}
    \left(
        \begin{array}{c}
            \varphi(0_+) \\
            \varphi(0_-)
        \end{array}
    \right)
\label{chi'}
\end{equation}
for some invertible constant matrix $\tilde{M}$. Here, we have
explicitly shown the energy dependence in the above relation.
Substituting eqs.$(\ref{chi})$ and $(\ref{chi'})$ into
eq.$(\ref{cc})$ leads to the connection condition for $\chi(x)$, i.e.
\begin{equation}
    (U-{\bf 1})\tilde{M}^{-1} \Phi'_{\chi}
     +iE L_0 (U+{\bf 1})M^{-1}\Phi_{\chi}=0.
\end{equation}
Since the connection condition must be independent of the energy $E$
and
since $\Phi_{\chi}$ and $\Phi'_{\chi}$ cannot vanish
simultaneously, we conclude that the eigenvalues of $U$
must be 
$\pm1$.\footnote{This situation resembles the discussion
of the Weyl scaling invariance \cite{symmetry}.
}
The case of $U=\pm {\bf 1}$ turns out to lead to no nontrivial models
because $\Phi_{\chi}$ and $\Phi'_{\chi}$ would vanish if we require
$\chi(x)$ to  satisfy the same connection condition as $\varphi(x)$.
The remaining possibility is that the two eigenvalues of $U$ are $+1$
and $-1$.
Then, the general form of $U$ is given by
\begin{equation}
  U = \exp\left\{i\frac{\pi}{2}({\bf 1}
      + \vec{\alpha}\cdot\vec{\sigma})\right\}
       \quad {\rm with} \ (\vec{\alpha})^2=1.
\end{equation}
This corresponds to the choice $(\theta_+, \theta_-)=(\pi,0)$ in
eq.(\ref{theta}). 
Thus, in terms of the eigenfunctions $\varphi_{\pm}$
of ${\cal G}=\vec{\alpha}\cdot \vec{{\cal P}}$, the connection
condition is reduced to
\begin{equation}
    {\rm type\ A}:~~ \varphi_+(0_+)=0=\varphi'_-(0_-).
	\label{a0}
\end{equation}
If we replace $\vec{\alpha}$ by $-\vec{\alpha}$, the role of
$\varphi_+$ and $\varphi_-$ is exchanged, so that we have another type
of allowed connection conditions.
\begin{equation}
    {\rm type\ B}: ~~ \varphi'_+(0_+)=0=\varphi_-(0_-).
\end{equation}
Repeating the same argument given above, we obtain two allowed connection
conditions at $x=l$, i.e.
\begin{eqnarray}
    {\rm type\ A}\!\!\!&:&~\varphi_+(l)=0=\varphi'_-(-l), \\
    {\rm type\ B}\!\!\!&:&~\varphi'_+(l)=0=\varphi_-(-l).
	\label{bl}
\end{eqnarray}
In the next section, we examine the models with the connection
conditions obtained above, and explicitly construct $N=2$ supercharges.
\section{Construction of $N=2$ supercharges}
In the previous section, we have found that compatibility with
supersymmetry restricts a class of connection conditions.
For every $su(2)$ element $g=\vec{\alpha}\cdot \vec{\sigma}$
(or ${\cal G}=\vec{\alpha}\cdot \vec{{\cal P}}$),
there are four types of the
allowed connection conditions; type (A,A), (B,B), (A,B) and (B,A).
(The first (second) entry denotes the type of
the connection condition at $x=0$ ($x=l$).)
In this section, we explicitly construct $N=2$
supercharges for the models, and show that the models of
type (A,B) and (B,A) possess no supersymmetric vacua,
so that supersymmetry is
spontaneously broken for those models.\footnote{
Our results for $g=\sigma_3$ reproduce those obtained
in ref.\cite{junker}.}

Let us first examine the type (A,A) model
whose connection conditions are given by
\begin{equation}
    \varphi_+(0_+)=\varphi'_-(0_-)=\varphi_+(l)=\varphi'_-(-l)=0.
\label{aa}
\end{equation}
Since ${\cal G}$ commutes with $H$, any energy eigenfunction
$\varphi_E(x)$ can be a simultaneous eigenfunction of ${\cal G}$. The
eigenfunctions are easily found as\footnote{
Without loss of generality, we can assume that $\alpha_3 \ne -1$.
The choice $\alpha_3=-1$ corresponds to ${\cal G}=-{\cal R}$.
This is physically equivalent to the choice $\alpha_3=1$ under the
exchange of $\varphi_+ \leftrightarrow \varphi_-$.
The reason why the 
expressions $(\ref{aaphi+})$ and $(\ref{aaphi-})$ become ill defined
for $\alpha_3=-1$ is that in this case $\varphi_{\pm}(x)=\Theta(\pm x)
\varphi(x)$, so that the domain of the function $\varphi_+(x)$
($\varphi_-(x)$) is given by $0<x<l$ ($-l<x<0$).
}
\begin{eqnarray}
    \varphi_{+,E_n}(x) &=&
     \Theta(x) A_n \sin{\left(\frac{n \pi}{l}x \right)}
      - \Theta(-x) A_n \frac{\alpha_1 + i \alpha_2}{1+ \alpha_3}
        \sin{\left( \frac{n \pi}{l}x\right)} ,
         \nonumber\\
             & &\qquad\qquad\qquad\qquad\qquad\qquad\qquad\qquad
             {\rm for} ~n=1,2,3,\cdots,
\label{aaphi+} \\
    \varphi_{-,E_n}(x) &=&
     -\Theta(x) A_n \frac{\alpha_1 -i\alpha_2}{1+\alpha_3}
      \cos{\left(\frac{n \pi}{l}x \right)}
       +\Theta(-x) A_n \cos{\left(\frac{n \pi}{l}x \right)} ,
         \nonumber\\
             & &\qquad\qquad\qquad\qquad\qquad\qquad\qquad\qquad
             {\rm for} ~n=0,1,2,\cdots,
\label{aaphi-}
\end{eqnarray}
where $A_n$ are normalization constants and the energy eigenvalues
$E_n$ are given by
\begin{equation}
    E_n =\frac{1}{2}\left( \frac{n\pi}{l} \right)^2
     \quad{\rm for} ~~n=0,1,2,\cdots.
\end{equation}
Thus, we found that the vacuum has vanishing energy and all the excited
states are doubly degenerate between even and odd ${\cal G}$-parity
states. To show the existence of supersymmetry, let us construct $N=2$
hermitian supercharges $Q_a~(a=1,2)$ satisfying
\begin{eqnarray}
    \{Q_a,Q_b \}&=&H\delta_{ab}, \label{salgebra1}\\
    (Q_a)^{\dagger} &=&Q_a.
\end{eqnarray}
It is important to note that the above relations are not
enough to guarantee $N=2$ supersymmetry.
We have to further require that for any state $\varphi(x)$ satisfying
the connection conditions (\ref{aa}),
the states $Q_a\varphi(x)$ ($a=1,2$) obey the same connection
conditions (\ref{aa}), 
otherwise $Q_a$ would not regarded as physical operators
of the system.

The connection conditions $(\ref{aa})$ and the fact that $Q_a$ are
proportional to $\frac{d}{dx}$ strongly suggest that the supercharges
$Q_a$ connect $\varphi_{+}$ and $\varphi_{-}$, i.e.
$Q_a\varphi_{\pm} \propto \varphi_{\mp}$. 
This implies that $Q_a$ should exchange the eigenvalues of ${\cal G}$
and hence anticommute with ${\cal G}$,
\begin{equation}
    Q_a {\cal G} =-{\cal G} Q_a ~~~{\rm for} ~a=1,2.
    \label{salgebra2}
\end{equation}
It follows that ${\cal G}$ can be regarded as the \lq\lq fermion" number
operator
\begin{equation}
    (-1)^F ={\cal G}.
\end{equation}
Noting that ${\cal P}_3 \frac{d}{dx}$ commutes with 
${\cal P}_j~(j=1,2,3)$, we can show that the following 
supercharges satisfy all the desired relations:
\begin{equation}
    Q_a =\frac{1}{2}{\cal G}_a {\cal P}_3 i \frac{d}{dx} , ~~~a=1,2,
  \label{Q}
\end{equation}
where
\begin{eqnarray}
    {\cal G}_a=\vec{\beta}_a \cdot \vec{{\cal P}} ~~~ {\rm
      with}\ (\vec{\beta}_{1})^2=(\vec{\beta}_{2})^2=1
      ~~{\rm and}  
       ~~\vec{\beta}_{1}\cdot \vec{\alpha}=\vec{\beta}_{2}\cdot
        \vec{\alpha}=\vec{\beta}_1 \cdot \vec{\beta}_{2}=0.
\end{eqnarray}
The vacuum state $\varphi_{-,E_0}(x)$ satisfies
$Q_a\varphi_{-,E_0}(x)=0$ for $a=1,2$, and hence supersymmetry is
unbroken.

Let us next consider the type (B,B) model whose connection conditions
are given by
$\varphi'_+(0_+)=\varphi_-(0_-)=\varphi'_+(l)=\varphi_-(-l)=0$.
The type (B,B) model turns out to be a dual theory of the type (A,A)
model because the transformation ${\cal G}_1$ (or ${\cal G}_2$)
gives a map from the Hilbert space of the type (A,A) model onto that of
the type (B,B) one due to the property
${\cal GG}_a =-{\cal G}_a{\cal G}$ \cite{symmetry}.
We will not discuss the type (B,B) model further.

Let us finally examine the type (A,B) model.
(The type (B,A) model is
physically equivalent to the type (A,B) model.)
The connection conditions of the model is given by
\begin{equation}
    \varphi_+(0_+)=\varphi'_-(0_-)=\varphi'_+(l)=\varphi_-(-l)=0.
\end{equation}
The energy eigenfunctions $\varphi_{\pm,E_n}(x)$ are found to be
\begin{eqnarray}
    \varphi_{+,E_n}(x) &=&
     \Theta(x) A_n \sin{\left(\frac{(n-\frac{1}{2}) \pi}{l}x \right)}
      - \Theta(-x) A_n \frac{\alpha_1 + i \alpha_2}{1+ \alpha_3}
        \sin{\left( \frac{(n-\frac{1}{2}) \pi}{l}x\right)} ,
         \nonumber\\
             & &\qquad\qquad\qquad\qquad\qquad\qquad\qquad\qquad\qquad
             {\rm for} ~n=1,2,3,\cdots, \\
    \varphi_{-,E_n}(x) &=&
     -\Theta(-x) A_n \frac{\alpha_1 -i\alpha_2}{1+\alpha_3}
      \cos{\left(\frac{(n-\frac{1}{2}) \pi}{l}x \right)}
       +\Theta(x) A_n \cos{\left(\frac{(n-\frac{1}{2}) \pi}{l}x \right)} ,
        \nonumber\\
             & &\qquad\qquad\qquad\qquad\qquad\qquad\qquad\qquad\qquad
             {\rm for} ~n=1,2,3,\cdots
\end{eqnarray}
with
\begin{equation}
    E_n= \frac{1}{2}\left( \frac{(n-\frac{1}{2})\pi}{l} \right)^2
      ~~~~{\rm for} ~n=1,2,3,\cdots.
\end{equation}
The supercharges $Q_a (a=1,2)$ are given by the same form as
eq.$(\ref{Q})$ and lead to the relations $Q_a \varphi_{\pm,E_n} \propto
\varphi_{\mp,E_n}$, as they should. All the energy eigenstates are
doubly degenerate and there is no 
vacuum state annihilated by $Q_a$, so
that supersymmetry is spontaneously broken in this model.
\section{Supersymmetric models with a superpotential}
In the previous sections, we have succeeded to construct the $N=2$
supersymmetric models only with point interactions.
In this section, we
extend the analyses to models containing a superpotential.
To this end, 
let us first recall that in the supersymmetric Witten model the
supercharges are given by \cite{witten}
\begin{eqnarray}
    Q_1^W&=& \frac{1}{2} \left[ \sigma_1 i \frac{d}{dx}+\sigma_2 W'(x)
\right], \\
    Q_2^W&=& \frac{1}{2} \left[ \sigma_2 i \frac{d}{dx}- \sigma_1 W'(x)
\right],
\end{eqnarray}
with the Hamiltonian
\begin{equation}
    H^W=\frac{1}{2} \left[ -\frac{d^2}{dx^2}+(W'(x))^2-\sigma_3 W''(x)
\right].
\end{equation}
A crucial observation in our formulation is that the set $\{{\cal G}_1
,{\cal G}_2,{\cal G}_3={\cal G} \}$ forms the $su(2)$ algebra of spin
$\frac{1}{2}$,\footnote{
The orthogonal unit vectors
$\{\vec{\beta}_1,\vec{\beta}_2 ,\vec{\alpha} \}$ are chosen such that
$\vec{\beta}_{1}\times \vec{\beta}_2 =\vec{\alpha}$.
}
i.e.
\begin{eqnarray}
    [{\cal G}_i,{\cal G}_j]&=&2 i \sum_{k=1}^{3}
    \epsilon_{ijk}{\cal G}_k ,\\
    \{{\cal G}_i,{\cal G}_j \}&=&2 \delta_{ij}.
\end{eqnarray}
Further, note that ${\cal P}_3 \frac{d}{dx}$ and ${\cal P}_3 W'(x)$
commute with ${\cal G}_j$ ($j=1,2,3$) if $W'(x)$ is an odd function,
i.e.
\begin{equation}
    W'(-x)=-W'(x).
\end{equation}
Although there is no such restriction on $W'(x)$ in the Witten
model, we require $W'(x)$ to be parity-odd in order for the
supercharges given below to satisfy the desired relations.
Since there are no reasons that the superpotential is smooth
at singularities, we allow $W'(x)$ to have discontinuity
there, so that $W'(0_{\pm})$ and $W'(\pm l)$ do not 
necessarily vanish.\footnote{%
We here assume that $W'(0_{\pm})$ and $W'(\pm l)$ are
finite to make the connection conditions 
(\ref{typea0})-(\ref{typebl}) well defined,
although some extension to allow divergent potentials
at singularities may be possible
(see ref.\cite{divergent}).
} 

The above observations may tell us to take supercharges to be of the form
\begin{eqnarray}
    Q_1 &=&\frac{1}{2} \left[ {\cal G}_1 {\cal P}_3 i \frac{d}{dx} +{\cal
G}_2 {\cal P}_3 W'(x) \right], \nonumber\\
    Q_2&=&\frac{1}{2} \left[ {\cal G}_2 {\cal P}_3 i \frac{d}{dx} -{\cal
G}_1 {\cal P}_3 W'(x) \right],
\label{scharge}
\end{eqnarray}
which satisfy the relations $(\ref{salgebra1})$ and $(\ref{salgebra2})$
with the Hamiltonian
\begin{equation}
    H=\frac{1}{2}\left[ -\frac{d^2}{dx^2} +(W'(x))^2 -{\cal G} W''(x)
\right] .
\label{hamiltonian}
\end{equation}
Again, the \lq\lq fermion" number operator is identified
with ${\cal G}$. 
The correspondence between the Witten model and our model is evident:
The Pauli matrices $\sigma_j$ in the Witten model are replaced by the
operators ${\cal G}_j$ in our model.
Both of them satisfy the $su(2)$ algebra of spin $\frac{1}{2}$.
Although physical meanings of ${\cal P}_3$ in front of $\frac{d}{dx}$
and $W'(x)$ are less obvious, it guarantees ${\cal P}_3 \frac{d}{dx}$
and $ {\cal P}_3 W'(x)$ to commute with ${\cal G}_j$.

We can further show that the supercharges become hermite 
and map
any state $\varphi_+$with ${\cal G} =+1$ onto some state $\varphi_-$
with ${\cal G}=-1$, and vice versa, if the connection conditions at
$x=0$ are chosen as
\begin{eqnarray}
    {\rm type ~A}&:&\varphi_+(0+)
      = 0 = \varphi'_-(0_-)-W'(0_-) \varphi_-(0_-),
	  \label{typea0}
\\
    {\rm type ~B}&:&\varphi'_+(0_+)+W'(0_+) \varphi_+(0_+)
      = 0 = \varphi_-(0_-), \label{typeb0}
\end{eqnarray}
and at $x=l$
\begin{eqnarray}
    {\rm type ~A}&:&\varphi_+(l)=0=\varphi'_-(-l)-W'(-l) \varphi_-(-l), 
	\label{typeal}\\
    {\rm type ~B}&:&\varphi'_+(l)+W'(l) \varphi_+(l)=0=\varphi_-(-l).
	\label{typebl}
\end{eqnarray}
Therefore, we have four types of the $N=2$ supersymmetric models for
every $su(2)$ element ${\cal G}$.
Let us note that the above connection conditions are reduced
to eqs.(\ref{a0})-(\ref{bl}) when $W'(x)=0$.

Let us finally discuss spontaneous breaking of supersymmetry for the
models obtained above.
The supersymmetric vacuum state is obtained by solving
\begin{equation}
Q_a \varphi_{\pm,0}(x)=0 ~~~~{\rm for} ~a=1,2.
\label{vacuum}
\end{equation}
Formal solutions to the above equations are given by
$\varphi_{\pm,0}(x) \propto{\rm exp}\{\mp W(x)\}$.
For a noncompact space, the normalizability of the states would remove
some of them from the Hilbert space.
However, since the space is
compact (a circle) in our model, the solutions to eq.$(\ref{vacuum})$
are always normalizable.
Nevertheless, some of them must be removed
from the Hilbert space.
For the type (A,A) ((B,B)) model, the state
$\varphi_{+,0}(x)$ ($\varphi_{-,0}(x)$) does not satisfy the connection
conditions and hence it does not belong to the Hilbert space of the
model.
On the other hand, $\varphi_{-,0}(x)$ ($\varphi_{+,0}(x)$)
satisfies the desired connection conditions
and it gives the supersymmetric vacuum.
Therefore, supersymmetry is unbroken in the
case of the type (A,A) and (B,B) models.
For the type (A,B) and (B,A) models, both $\varphi_{\pm,0}(x)$ do not
satisfy the connection conditions at $x=0$ or $x=l$. Hence
supersymmetry is spontaneously broken in these models.\footnote{
Other mechanisms of (spontaneous) supersymmetry breaking due to
boundary effects have been found in ref.\cite{susybreaking1,
susybreaking2}.}

\section{Summary and discussions}
In this paper, we have investigated quantum mechanics on a circle
with point interactions and clarified a class of connection conditions 
compatible with supersymmetry.
The representation of the constructed $N=2$ supercharges
turns out to reflect the characteristics  of quantum
mechanics with point singularities because the supercharges are
represented in terms of the discrete
transformations ${\cal G}_j$ ($j=1,2,3$), 
which make wavefunctions discontinuous in general and hence
are meaningless in ordinary quantum theory with no singularities.

It is interesting to note that for a special case of 
${\cal G}={\cal P}$ with any smooth odd function $W'(x)$,
the connection conditions for the type (B,B) model
are reduced to the conditions that wavefunctions must
be smoothly connected at $x=0$ and $x=l$.
In other words, this model has no singularity at all.
In this case, the form of the Hamiltonian (\ref{hamiltonian})
and the supercharges (\ref{scharge}) with 
${\cal G}_1={\cal Q}$ and ${\cal G}_2={\cal R}$ has been
found in ref.\cite{bosonization},
as a minimal bosonization of $N=2$ supersymmetry.
Our results may be considered as a natural extension of
the minimal bosonization of $N=2$ supersymmetry with point
singularities.

In ref.\cite{susy}, supersymmetry in the system
of a free particle on a line ${\bf R}$ or an interval $[-l, l]$
with a point singularity at $x=0$ was discussed.
Although the configuration spaces are slightly different each
other, some of the results overlap with ours.
To see this, let us restrict the connection conditions
in our model to a specific class of $g=\sigma_{3}$
(or ${\cal G}={\cal R}$).
Then, a point singularity associated with $g=\sigma_{3}$
describes a perfect wall through which no probability
flow can penetrate, so that the circle $S^1$ can
be regarded as an interval $[-l, l]$ with a point singularity
at $x=0$ in ref.\cite{susy}.
We further need to restrict the superpotential to be constant
($W'(x) = - W'(-x) = b$ for $0<x<l$), in order to have
the free Hamiltonian.
According to the prescription of ref.\cite{susy}, 
we introduce a two-component wavefunction 
$\Psi(x) = (\psi_{+}(x), \psi_{-}(x))^T$ for $x>0$, where
$\psi_{+}(x) \equiv \varphi(x)$ for $x>0$ and
$\psi_{-}(-x) \equiv \varphi(x)$ for $x<0$.
Noting that $\psi_{\pm}(x)$ are the eigenfunctions of
${\cal R}$ with ${\cal R} = \pm 1$, respectively,
and taking ${\cal G}_1 = {\cal P}$ and 
${\cal G}_2 = {\cal Q}$, we find that the action of
$Q_a (a=1,2)$ in eqs.(\ref{scharge}) on $\Psi (x)$
can be represented by the $2\times 2$ matrices:
\begin{eqnarray}
  Q_1:\Psi(x)\  \rightarrow \ 
    \tilde{Q}_1 \Psi(x) &=& 
	  \left( \frac{i}{2} \frac{d}{dx}\sigma_1 +
	    \frac{b}{2}\sigma_2\right) \Psi(x) , \nonumber\\
  Q_2:\Psi(x)\  \rightarrow \ 
    \tilde{Q}_2 \Psi(x) &=& 
	  \left( \frac{i}{2} \frac{d}{dx}\sigma_2 -
	    \frac{b}{2}\sigma_1\right) \Psi(x) .
\label{2x2matrixQ}
\end{eqnarray}
The above representation of the supercharges agrees with that of
ref.\cite{susy} (see eqs.(2.21)), up to normalization.
The allowed connection conditions (\ref{typea0})-(\ref{typebl}) 
are also found in ref.\cite{susy}
(see eqs.(3.17) and (3.20)).\footnote{
We, however, missed $N=1$ supersymmetry found in ref.\cite{susy}.
This is due to the fact that the connection conditions at
$x=0$ and $x=l$ are taken to be the same class associated
with an $su(2)$ element $g$ in our model.
}
Thus, our results give an extension of the work \cite{susy}
for a free particle to a supersymmetric system with a
superpotential.
It should be stressed that although the supercharges
(\ref{scharge}) could be represented by $2\times 2$ matrices,
as done above for a special case, 
our representation of the supercharges has the advantage of
clarifying
the role of the discrete transformations ${\cal P}_j$.
Our approach will be useful in analyzing more complex
systems.

Our success of obtaining $N=2$ supersymmetry may lead to an
expectation to have higher $N$-extended supersymmetry by putting
a number of point interactions on a circle.
This turns out to be true but a class of allowed connection
conditions compatible with higher $N$-extended supersymmetry
is more restrictive \cite{N-SUSY}.
A simplest example of $N=4$ supersymmetry
without a superpotential can be obtained
by putting four point singularities at $x=0, \pm l/2, l$
on a circle and by choosing all the connection conditions
to be associated with $g=\sigma_3$ and the type A.
Then, we find that the doubly degenerate vacua have a vanishing
energy and that all the excited states are four-fold degenerate.
The degeneracy results from $N=4$ supersymmetry and
we can construct four hermite supercharges.
An easy way to represent the supercharges is to introduce,
as a natural extension of the work \cite{susy},
a four-component wavefunction
$\Psi(x) = (\psi_{1}(x), \psi_{2}(x), 
\psi_{3}(x), \psi_{4}(x))^T$ for $0<x<l/2$, where
$\psi_{1}(x) \equiv \varphi(x)$,
$\psi_{2}(x) \equiv \varphi(-x)$,
$\psi_{3}(x) \equiv \varphi(l-x)$,
and
$\psi_{4}(x) \equiv \varphi(-l+x)$ for $0<x<l/2$.
In this basis, the supercharges can be represented by
the $4\times4$ matrices such as
\begin{eqnarray}
  \tilde{Q}_1 &=& 
	  \frac{i}{2} \frac{d}{dx}
	    \left( 
		  \begin{array}{cc}
		    \sigma_1 & 0 \\
			0 & \sigma_1
		  \end{array}
        \right) , \qquad
  \tilde{Q}_2 = 
	  \frac{i}{2} \frac{d}{dx}
	    \left( 
		  \begin{array}{cc}
		    \sigma_2 & 0 \\
			0 & \sigma_2
		  \end{array}
        \right) , \nonumber\\
  \tilde{Q}_3 &=& 
	  \frac{i}{2} \frac{d}{dx}
	    \left( 
		  \begin{array}{cc}
		    0 &\sigma_3 \\
			\sigma_3 & 0
		  \end{array}
        \right) , \qquad
  \tilde{Q}_4 =
	  \frac{i}{2} \frac{d}{dx}
	    \left( 
		  \begin{array}{cc}
		    0 & -i\sigma_3 \\
			i\sigma_3 & 0
		  \end{array}
        \right) .
\label{4x4matrixQ}		
\end{eqnarray}
The extension to higher $N$-extended supersymmetry and the
inclusion of a superpotential are possible.
The subject will be reported elsewhere \cite{N-SUSY}.

\section*{Acknowledgements}
The authors wish to thank
T. Kobayashi,
M. Kubo,
T. Kugo,
C.S. Lim,
I. Tsutsui
for valuable discussions and useful comments. K.T. would like to thank
the Dublin Institute for Advanced Studies for warm hospitality.

\baselineskip 5mm 

\end{document}